\documentstyle[prd,aps,floats,epsf]{revtex}
\newcommand{\be}{\begin{equation}}
\newcommand{\ee}{\end{equation}}
\newcommand{\G}{\Gamma}

\begin{document}
\preprint{PURD-TH-XXYY, hep-ph/yymmddd}
\draft

\renewcommand{\topfraction}{0.99}
\renewcommand{\bottomfraction}{0.99}
\twocolumn
%[\hsize\textwidth\columnwidth\hsize\csname@twocolumnfalse\endcsname]

\title
{\Large 
%{\bf On the Influence of Isocurvature Perturbations in
%Quintessential Models}
{\bf Attractors and Isocurvature Perturbations in Quintessence Models} }
\author{L. R. Abramo$^{\, 1}$ and F. Finelli$^{\, 2,3}$ } 
\address{$^1$ Ludwig-Maximilians Universit\"at, Sektion Physik, \\
Theresienstr. 37, 80333 Munich, Germany \\
$^2$ Department of Physics, Purdue University, West Lafayette, IN 47907,
USA \\
$^3$Dipartimento di Fisica, Universit\`a degli Studi di Bologna
and I.N.F.N., \\ via Irnerio, 46 -- 40126 Bologna -- Italy
}
\date{\today}
\maketitle

\begin{abstract}
We investigate the evolution of cosmological perturbations in  
scenarios with a quintessence scalar field, both analytically and 
numerically. In the tracking regime for quintessence, 
we find the long wavelength solutions for the perturbations of the
quintessence field. 
We discuss the possibility of isocurvature modes generated by the 
quintessence sector and their impact on observations.
\end{abstract}

\vskip -0.5cm

\vskip 0.4cm
 
\section{Introduction}

Observational data seems to indicate that we live in an
accelerating Universe \cite{supernova}. As an
alternative to the scenario where the acceleration is fueled 
by a cosmological constant $\Lambda$, models with a scalar field 
capable of dominating recently and of developing a negative pressure 
have been proposed \cite{ZWS}. This scalar field which pervades our 
universe has been dubbed {\em quintessence}.

Compared to the cosmological constant, quintessence has two 
important differences. First, quintessence can be interpreted 
as a fluid with a time dependent equation of state. Therefore quintessence 
models may alleviate the so-called coincidence problem,
which is the apparent cosmic collusion that  
the dark energy component is fine-tuned in a way that 
it is starting to dominate 
the energy density of the universe just at the present time.
And second, in contrast to the cosmological {\em constant}, 
the quintessence field can fluctuate \cite{ZWS,RP,FJ,VL,PB,BMR}.

One interesting possibility is that the quintessence field 
$Q(\vec{x},t)$ can, in combination with the other cosmic fluids 
(radiation, baryons, cold dark matter, etc.), lead 
not only to adiabatic (curvature) perturbations, but to a mixture 
which includes an isocurvature component.
Isocurvature (or entropy) perturbations appear when the relative 
energy density and pressure perturbations of the different fluid 
species combine to leave the overall curvature perturbations 
unchanged. In quintessence models, the presence 
of potentially relevant isocurvature 
modes could be generic, just as in multi-field 
inflationary models \cite{MULTI}. 
Indeed, quintessence is constructed is such a way that
it is an unthermalized component subdominant for most of the history of
the Universe. Since quintessence is uncoupled from the rest of matter
because of astrophysical and cosmological constraints 
\cite{constraints}, its 
fluctuations may lead to an isocurvature component, whose nature 
will be preserved except for the known integrated feeding of the
adiabatic component, expressed by the relation \cite{MFB,G-BW}:
\be
\dot \zeta = \frac{2}{3 H (1+w)} \left[ c_s^2 \frac{\nabla^2
\Phi}{a^2} + \frac{1}{2} \delta p_{\rm nad} \right] \; ,
\label{zeta}
\ee
where $\zeta$ is the gauge-invariant curvature perturbation,
a dot denotes time derivative, 
$w \equiv p/\rho$ and $c_s^2 \equiv \dot{p}/\dot{\rho}$ denote 
respectively the equation of state and the
speed of sound of the total matter content,
$p_{\rm nad}$ is the total non-adiabatic pressure perturbation and we
employ units such that $8 \pi G = 1$.

This latter effect was studied \cite{ABT} in the context of axion
perturbations, and \cite{MOLLERACH} in the context of the baryon
isocurvature model \cite{PEEBLESOLD}. This effect is also responsible for
the growth of super-Hubble adiabatic perturbations during preheating
\cite{FB}. Considering axions as Cold Dark Matter (CDM) \cite{ABT}, an
isocurvature perturbation due to the angle misalignment produced
during inflation induces an adiabatic component
of comparable amplitude at the moment of reentry of the perturbation
inside the Hubble radius. This is due to the fact that the CDM component 
is going to dominate about the time of decoupling, and thus 
the integrated effect is almost completed by the time that
mode reenters inside the Hubble radius.

However, the quintessence case is different from the axion/CDM 
since in most of the models quintessence fluctuations are damped
inside the Hubble radius. This is required in order to minimize the 
impact of an additional dynamical degree of freedom on structure 
formation. The quintessence and the axion/CDM differ also in 
either one of two ways: i) Q was still a negligible component before 
the time of decoupling, or 
ii) Q was comparable to normal matter, $\Omega_Q = {\cal{O}}(1)$, but 
in a so-called {\em tracking} regime \cite{FJ,SWZ} whereby its 
equation of state $w_Q \equiv p_Q/\rho_Q$
was approximately that of dust or radiation --- 
whichever was dominating at the time. In the first case (which happens, 
for example, in the PNGB scenario \cite{PNGB}) isocurvature perturbations 
are irrelevant simply because the field $Q(\vec{x},t)$ 
is a negligible component until a redshift of at least $z \sim 10$.
In the second case the energy density in $Q$ need not be small, however
due to the tracking of the quintessence field,
perturbations in the $Q$-fluid behaved similarly to the perturbations 
in the background for most of the observable history of the Universe, 
and isocurvature perturbations are therefore suppressed until
the end of the tracking phase.
In either case, a primordial isocurvature perturbation could still 
be present, but it would not have had enough time to induce an 
adiabatic component.

From an observational point of view, isocurvature perturbations have a
very distinctive imprint on the spectrum of the temperature anisotropy of
the Cosmic Microwave Background (CMB) 
\cite{EB,BKT,LR}. 
With the accuracy that 
future CMB experiments such as MAP \cite{MAP} and Planck \cite{PLANCK}
will be able to reach, 
the constraints
on the ratio of uncorrelated isocurvature perturbations in CDM-radiation 
to the adiabatic component will be of 
the order of percents \cite{PIER,ENKV}. 
Recently the impact of generic isocurvature modes on the estimation of 
cosmological parameters has been also investigated \cite{BMT}.
Therefore it is important to 
understand the evolution of isocurvature modes in quintessence models 
where an unthermalized, uncoupled relic survives
until the present era, and dominates very recently.
%The primordial spectrum of quintessence perturbations,
%produced in quintessential models at decoupling, 
In most of the literature a primordial adiabatic spectrum for quintessence
perturbations is assumed. 
If the notion of adiabaticity among different components is related to
their thermal equilibrium, then the weakly coupled nature of quintessence
could evade this condition. The primordial spectrum could be generated
during inflation and/or influenced through its evolution until the
decoupling time.

The outline of the paper is as follows. In section II we give necessary
notions of the background evolution of quintessence models. In section III 
we study the evolution of cosmological perturbations and we identify the
attractor solution for quintessence perturbations during the tracking
regime. In section IV we study the evolution of isocurvature perturbations
and their feedback on the adiabatic component. In section V we discuss
the initial condition for quintessence perturbations after
nucleosynthesys. We conclude in section VI.

\section{Background evolution with quintessence}

For simplicity we consider only radiation, pressureless matter and 
quintessence, and ignore the neutrinos as well as the distinction 
between baryons and CDM. Each component
$i$ ($i = r,m,Q$) has an energy density $\rho_i$ and pressure $p_i$. 
The sum of the energy densities determines the Hubble parameter via the usual
Einstein equation, $3H^2 = \rho = \sum_i \rho_i $. The equation
of state $w_i \equiv p_i/\rho_i$ is $1/3$ and zero in the case of radiation
and matter, respectively. The background energy density and
the pressure of the quintessence field are:
\be
\label{rhoQ}
\rho_Q = \frac12 \dot{Q}^2 + V(Q) \quad , \quad p_Q = 
\frac12 \dot{Q}^2 - V(Q) \; ,
\ee
where the background scalar field $Q(t)$ obeys the equation 
$\ddot{Q} + 3H \dot{Q} + V_{,Q}=0$. 
This means that $w_Q$ is in general time-dependent.

The main requirement of quintessence is that it starts to dominate
the energy density of the Universe only at the present time, with 
an equation of state $w_Q(z=0) \equiv w_{Q0} < 0$. Conservative 
phenomenology dictates that $\rho_{Q0}/\rho_0 = \Omega_{Q0} < 0.8$ 
--- to allow time 
for galaxy formation --- and that $w_{Q0} < -0.5$ --- to accommodate 
the SNIa data \cite{Pheno}. In addition, nucleosynthesis demands
that $\Omega_Q < 0.2$ at $z \simeq 10^9$.

Model building should obtain these values and still manage to solve
(or at least alleviate) the coincidence problem without too much 
fine tuning (see, e.g., \cite{Kess}.) 
The main problem seems to be that in order to
solve the coincidence problem one needs a period of tracking, but it is
hard to get a period of tracking and still obtain $w_{Q0} < -0.5$.
It is certainly possible to construct potentials which implement both
conditions \cite{BMR}, however we prefer to look at simpler potentials
which contain features which are generic to most of the models.
As it turns out, the relevant features of the cosmological perturbations 
do not depend on the specific
form of the potential, but only on its generic phenomenology.

In a typical scenario, nicely reviewed in \cite{BMR},
the scalar field starts out subdominant deep in the radiation era, in a
{\em kinetic} phase with an equation of state $w_Q = + 1$ (see Fig. 1.)
The kinetic energy $\frac12 \dot{Q}^2$ eventually decays, leaving only the 
quintessence potential energy, which is nearly constant and as a result 
$w_Q \rightarrow -1$ (of course, the quintessence field could also start
already in the potential-energy dominated regime). This is the so
called {\em potential} phase.
When $\Omega_Q$ becomes of order unity, the quintessence field
undergoes a transition which puts it into a {\em tracking} regime, where
it follows approximately the equation of state of the background.
Finally, at some point late in the matter era $Q$ starts to dominate
and the Universe begins the accelerated expansion phase that we observe 
today (the Q-dominated phase.)

\begin{figure}
\centerline{\epsfxsize=0.5\textwidth\epsffile{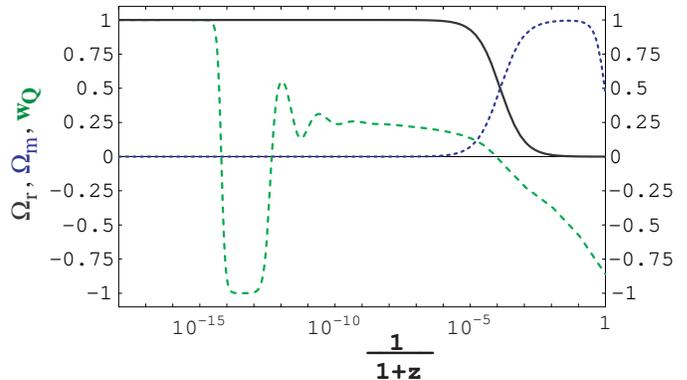}}
\caption{Densities of radiation (solid line, black) and matter (dotted
line, blue), and the equation of state for the scalar field (dashed line,
green), as a function of redshift for a model with potential $V(Q) = M^4 
e^{f/Q}$. In this plot $M^4 = 10^{-70} M_{pl}^4$  and $f=1 \, M_{pl}$
(with $8 \pi G = M_{pl}^{-2}$.)}
\label{fig1}
\end{figure}

As long as we keep away from the time of equal matter and radiation 
($z_{eq} \simeq 10^4$ in our flat models with $H_0 = 65$ Km s$^{-1}$ 
Mpc$^{-1}$), one of the two barotropic fluids (radiation or matter) 
can be neglected. 
The equation of state of the total matter content then reads:
\be
\label{wOmega}
w \equiv \frac{p}{\rho} = 
w_F + (w_Q-w_F) \Omega_Q \; ,
\ee
where the subscript $F$ stands for either radiation, 
when $t \ll t_{eq}$, or matter, when $t \gg t_{eq}$.
The total speed of sound $c_s^2$ has a simple expression as well:
\be
\label{cs2Om}
c_s^2 \equiv \frac{\dot p}{\dot \rho}
= w_F + \frac{\dot{Q}^2}{\rho+p} (c_Q^2 - w_F) \; ,
\ee
where $c_Q^2 \equiv \dot{p}_Q / \dot{\rho}_Q$. It is useful to note that
\be
\label{wQdot}
c_Q^2 = 1 + \frac{2 V_{,Q}}{3H\dot{Q}} = 
w_Q - \frac{\dot{w}_Q}{3H(1+w_Q)} \; .
\ee
When $c_Q^2=$constant, then $\dot Q \propto V_{,Q}/H$.
Taking the time derivative of $c_Q^2$ we also obtain:
\be
\label{VQQ}
\frac{2 V_{,QQ}}{3H^2} = \frac{1}{H} \frac{d}{dt} c_Q^2 +
(c_Q^2 - 1) \left[ 1+\frac{\dot{H}}{H^2} 
- \frac12 \left( 3 c_Q^2 + 5 \right) \right] \; .
\ee
When $w_Q$ is approximately constant, then from Eq. (\ref{wQdot})
$c_Q^2$ is constant as well. 
By using the background relation $\dot{H}/H^2 \simeq -3 (1+w_F)/2 $ 
we obtain
\be
V_{,QQ} \simeq \frac{9}{4} H^2 (1 - c_Q^2) (w_F + c_Q^2 + 2) \equiv \alpha
H^2 \; .
\label{VQQ2}
\ee
We observe that quintessence fluctuations are effectively massless 
($V_{QQ} \simeq 0$) during the kinetic phase ($c_Q^2 =1$) and the 
potential phase ($c_Q^2 = -2 - w_F$ \cite{BMR}).

A good quantity to measure how closely the quintessence field $Q$ tracks 
the background is the quantity $\gamma \equiv V_{,QQ} V/(V^2_{,Q})$ --- 
which was named $\G$ in \cite{ZWS}. When $w_Q$ is approximately
constant in time, then:
\be
\label{gamma_appr}
\gamma \simeq 1 + \frac{w_F - w_Q}{2 (1+w_Q)} \,.
\ee
If tracking is {\em exact} (as is the case in the exponential potential
models), $w_F = w_Q$, then $\gamma=1$ and $V_{,QQ}/V_{,Q} = V_{,Q}/V$.

If $w_Q$ is approximately constant, then the kinetic 
and potential energies of the scalar field $Q$ must be proportional to 
each other. Taking the time derivative of 
$\rho_Q \propto \dot{Q}^2 \propto V$ we obtain that during a 
tracking phase:
\be
\label{VpV}
\frac{V_{,Q}}{V} 
= - 3 (1+w_Q) \frac{H}{\dot{Q}} \simeq {\rm constant} \; ,
\ee
where we have used the fact that $\dot{Q}^2 \propto a^{-3(1+w_Q)}$ to
find $H/\dot{Q} \propto a^{-3(w_F-w_Q)} \propto$ constant, if 
$w_Q \simeq w_F$.

We note that during the phases in which $w_Q$ is constant $\Omega_Q$ goes
as: 
\be
\Omega_Q \equiv \frac{\rho_Q}{\rho_{\rm TOT}} = a^{3(w_{\rm TOT} - w_Q)}
\,.
\ee
Therefore in the radiation epoch $\Omega_Q$ redshifts as 
$a^{-2}$ during its kinetic phase, and it grows as $a^4$ 
during its potential phase. During
the tracking regime $\Omega_Q$ is approximately 
constant (depending on how accurate is the tracking) 
and in the Q-dominated regime it is evidently constant (since 
$\Omega_Q \rightarrow {\cal O}(1)$ when $Q$ dominates.)

\section{Evolution of perturbations}

We compute the evolution of the cosmological perturbations in
longitudinal (or conformal-Newtonian) gauge \cite{MFB}:
\be
ds^2 = (1 + 2\Phi) dt^2 - a^2(t) (1 - 2 \Phi) d{\bf x}^2 \; .
\ee
In the matter sector,
the perturbed energy density and pressure of quintessence are,
as usual,
\begin{eqnarray}
\label{drQ}
\delta\rho_Q &=& 
\dot{Q} \delta \dot{Q} - \dot{Q}^2 \Phi + V_{,Q} \delta Q \; ,
\\ 
\label{dpQ}
\delta p_Q &=& 
\dot{Q} \delta \dot{Q} - \dot{Q}^2 \Phi - V_{,Q} \delta Q \; ,
\end{eqnarray}
where the Fourier transforms of the scalar field fluctuations obeys
the equation:
\be
\delta \ddot{Q} + 3 H \delta \dot{Q}
+ \frac{k^2}{a^2} \delta Q + V_{,QQ} \delta Q  
= + 4 \dot{Q} \dot\Phi - 2 V_{,Q} \Phi \; .
\label{fieldpert}
\ee 
The density fluctuations in radiation and matter, on the other hand,
obey the conservation equations for the density contrasts
$\delta_F \equiv \delta \rho_F/\rho_F$:
\be
\label{consE}
\dot{\delta}_F - 3 (1+w_F) \dot{\Phi} = (1+w_F) \frac{k}{a(t)} V_F \; ,
\ee
where $V_F$ is the fluid velocity. In the long wavelength limit 
($k \ll a H$) this equation is extremely useful, since it reads:
\be
\label{consE2}
\delta_F - 3(1+w_F)\Phi =  {\rm constant} = \delta_F^i - 3 (1+w_F)\Phi^i
\; ,
\ee
where the superscript $i$ indicates that the fluctuations have been
evaluated at some initial time $t_i$.
We can also combine Eq. (\ref{consE}) 
for $\delta_F$ with the equation for conservation of momentum,
\be
\label{V2}
\dot{V}_F = - \frac{k}{a(t)} 
\left( \frac{w_F}{1+w_F} \delta_F + \Phi \right) \; ,
\ee
and obtain a simple second order equation for $\delta_F$:
\be
\label{ConsEq}
\delta_F \, '' + w_F k^2 \delta_F = 
3 (1+w_f) \left( \Phi \, '' - \frac{k^2}{3} \Phi \right) \; ,
\ee
where a prime denotes, as usual, derivative with respect to conformal
time $\eta$, where $d/d\eta \equiv a(t) d/dt$.

We consider now the evolution of the quintessence field perturbations. 
At first, let us neglect the metric perturbations, i.e. take 
Eq. (\ref{fieldpert}) without its
right-hand side. By using the rescaled variables
$\delta \tilde{Q} = a^{1/2} \delta Q$,
Eq. (\ref{fieldpert}) can be rewritten as:
\be
\delta \tilde{Q} '' + {\cal H} \delta \tilde{Q} '
+\left[ k^2  + a^2 V_{, QQ} - \frac{a''}{2 a} - \frac{ {\cal
H}^2 }{4} \right] \delta \tilde{Q} = 0 \; ,
\label{fieldpert2}
\ee
where ${\cal{H}}= a'/a = a H$.

In a radiation dominated universe $a \propto \eta$ and
the term proportional to the second time derivative of the scale
factor vanishes; when Eq. (\ref{VQQ2}) holds, then the solutions
for the field perturbations in rigid space-times are:
\begin{displaymath}
%\label{fixedsolution}
\delta Q \sim \eta^{-1/2} \times \left\{
\begin{array}{c}
J_{|\nu|} (k \eta) \\
J_{-|\nu|} (k \eta)
%\label{fixedsolution}
\end{array}
\right. \; ,
%\label{fixedsolution}
\end{displaymath}
where
\be
\nu^2 = \frac14 - \alpha \,.
\label{nu}
\ee
If $\alpha \ne 0$ both the solutions decay in time. 
If $\alpha \rightarrow 0$ then there is a constant mode.

The argument above holds for a matter dominated universe as well: 
if $\alpha=0$ there is a constant mode, otherwise both the solutions 
decay in time (for a matter dominated universe the appropriate 
rescaled variable is $\delta \tilde{Q} = a^{3/4} \delta Q$.)

The inclusion of gravitational fluctuations in Eq. (\ref{fieldpert}) 
leads to a constant solution for the quintessence perturbations in the
long-wavelength limit. From Eq. (\ref{fieldpert}) we immediately
see that in this limit there are {\em constant} 
solutions $\Phi(t) \rightarrow \Phi^c$ and 
$\delta Q(t) \rightarrow \delta Q^c$:
\be
\label{ndec}
\delta Q^c \simeq - 2 \frac{V_{,Q}}{V_{,QQ}} \Phi^c \; ,
\ee
as long as $V_{,Q}/V_{,QQ}$ is approximately constant.
But this is precisely what happens during the tracking regime: from
Eqs. (\ref{VpV}) and (\ref{gamma_appr}) 
we see that $V_{,Q}/V_{,QQ} \simeq V/V_{,Q} \simeq$ 
constant in the tracking period. 
We stress that the type of solution (\ref{ndec}) does not hold
in the kinetic and potential phases, since in these cases 
$V_{,QQ} \rightarrow 0$.

We can use the $0-0$ component of the Einstein equations to relate the
Newtonian potential to the energy densities of other fluids
\be
\label{EFE}
-6 H^2 \Phi - 6 H \dot\Phi - 2 \frac{k^2}{a^2} \Phi = 
\delta \rho_r + \delta \rho_m  + \delta \rho_Q \; .
\ee
In the long wavelength limit, assuming that $\Phi$ is stationary 
and ignoring the subdominant barotropic fluid we obtain:
\be
\label{EFEapp}
- 2 \Phi \simeq \Omega_F \delta_F + \Omega_Q \delta_Q\; .
\ee 

In the regime described by the attractor in Eq. (\ref{ndec}) the perturbed
energy density for quintessence, defined in Eq. (\ref{drQ}), reduces to: 
\begin{eqnarray}
\delta \rho_Q^c &\simeq & -2 \left( 1 - \frac{V_{,Q} V_{,QQQ}}{V_{,QQ}^2} \right)
\dot{Q}^2 \Phi^c 
- 2 \Phi^c ( \frac{\dot{Q}^2}{2} + \frac{V}{\gamma}  ) \nonumber
\\ \label{drhoc}
&\simeq& - 2 \, \Phi^c \left[ \rho_Q 
- ( \dot{Q}^2 + V) \frac{\gamma - 1}{\gamma} + 
{\cal O}(\dot \gamma) \right]
\; .
\end{eqnarray}
and the perturbed pressure defined in Eq. (\ref{dpQ}) is:
\begin{equation}
\label{dpress}
\delta p_Q = - 2 \, p_Q \, \Phi
+ 2 \Phi ( \dot{Q}^2 - V) \frac{\gamma - 1}{\gamma} + {\cal O}(\dot \gamma)
\; .
\end{equation}
%We note that if tracking is exact we 
%have $\delta_Q = \delta_F = - 2 \Phi $. 
Using now the background identities 
$3H^2 = \rho$ and $\Omega_Q \simeq 1 - \Omega_F$ into Eq. 
(\ref{drhoc}), we obtain the density contrasts as functions of the 
Newtonian potential in the tracking regime:
\be
\label{Phirho}
\delta_Q^c \simeq \delta_F^c \simeq - 2 \Phi^c \; .
\ee
Notice that during tracking, quintessence and the dominant fluid 
species are in effect indistinguishable 
($\delta_Q^c \simeq \delta_F^c$), consequently we expect 
isocurvature perturbations to be suppressed during that period.

A solution corresponding to (\ref{ndec}) --- 
though in the synchronous gauge --- was first obtained in the 
case of exponential potentials, for which tracking is exact \cite{FJ}.
An attractor for quintessence perturbations has also been conjectured 
in \cite{BMR}. As we have shown, the approximate solutions (\ref{ndec}) 
and (\ref{Phirho}) hold for any potential, just as long as there is 
tracking, and, of course, they exist in any gauge.

We can also combine Eqs. (\ref{Phirho}) and (\ref{consE2}) to 
obtain a relationship between the initial and final values of the
Newtonian potential:
\be
\label{INICO}
\Phi^c \simeq 
- \frac16 \left( \delta_r^i - 4 \Phi^i \right) \; ,
\ee
where $\delta_r^i$ and $\Phi^i$ are initial conditions for
the radiation density contrast and the Newtonian potential respectively.
Notice that these initial conditions can be specified even at a 
time when quintessence is dominating, and even if the radiation contrast
and the Newtonian potential are initially not constant.

We have numerically verified formulas (\ref{ndec}),
(\ref{drhoc}), (\ref{Phirho}) and (\ref{INICO}) for 
several scenarios and initial conditions. 
Take for example the scenario whose background appeared in our Fig. 1.
Two typical initial conditions for the field perturbations are plotted 
in Figs. 2A and 2B (solid and dashed lines), together with the
attractor solutions (thin red lines): 
as the scalar field starts to approach the tracking regime at 
$z \simeq 10^{12}$, the perturbations start to converge around the
attractor solution. As seen in Fig. 2A for the field perturbations
$\delta Q$, the attractor of Eq. (\ref{ndec})
is a very good approximation even after the tracking 
phase has ended --- that is, Eq. (\ref{ndec}) is a very good approximation 
during the quintessence-domination period as well, even though $\delta Q$ 
is not constant anymore.
Fig. 2B shows how the quintessence density contrasts 
converge to $-2 \Phi$. 
In fact, for a wide range of initial values
the scalar field perturbations end up at the same solution 
$\delta Q^c$ after tracking.

\begin{figure}
%  \centerline{\epsfxsize=0.45\textwidth\epsffile{Attractor_ExpInvQ.eps}}
  \centerline{\epsfxsize=0.45\textwidth\epsffile{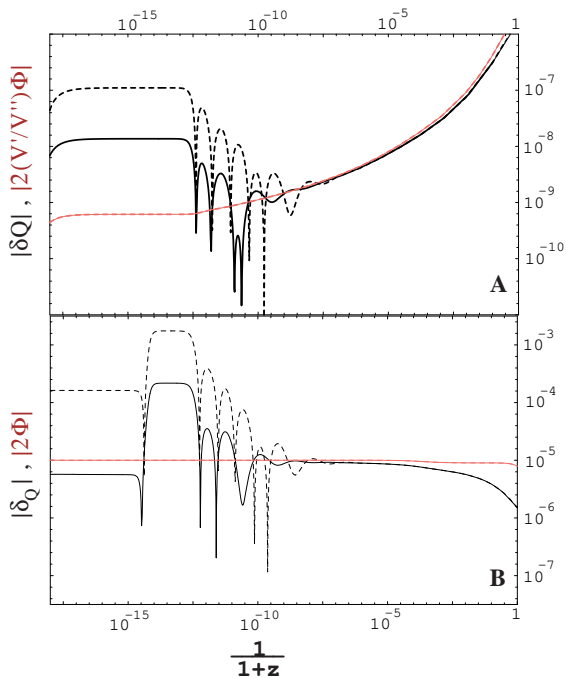}}
  \caption{Quintessence field fluctuations $\delta Q$ (panel A) and 
quintessence energy density contrasts $\delta_Q$ (panel B) for two 
different initial 
conditions for $\delta Q$ and $\delta \dot{Q}$ (solid and dashed lines.) 
We have chosen initial conditions such that
$\Phi \simeq 10^{-5}$ at $z_{\rm dec}=10^3$.
In panel A the quintessence fluctuations (dashed and solid black lines) 
are compared with the attractor $- 2 \Phi V_{,Q}/V_{,QQ}$ (thin red line).
Notice that even after the tracking phase ends 
the analytic approximation of Eq. (\ref{ndec}) remains very good.
In panel B the quintessence energy density contrasts (solid and dashed 
black lines) are compared 
with $-2 \Phi$ (thin red line), verifying the second approximation, 
Eq. (\ref{Phirho}). The approximation becomes worse as the tracking 
ends. In both plots the wavelengths of the modes cross 
the Hubble radius at $z=0$.} 
\label{fig2}
\end{figure}

This can be also seen in Fig. 3, 
which is the phase diagram for the perturbations with
different initial conditions shown in Fig. 2. 
During tracking the solutions spiral down to the attractor
(solid and dashed lines.)
When tracking ends the attractor disappears, but by that time most
modes have settled down to the same value, and their evolution is
henceforth the same (see the thin red line in Fig. 3 which springs
from the attractor point.)
We note that this attractor occurs generally only 
during tracking and for long wavelengths: when the wavelength 
becomes important (of the order of the inverse effective mass 
\footnote{If the effective mass is of the order of the
Hubble radius, this occurs when the perturbations 
reenter inside the Hubble radius. It occurs at the same time if
the effective mass vanishes (as happens during the kinetic and potential
phases).}), there is a sensitivity 
with respect to the initial conditions of the 
quintessence perturbations.

\begin{figure}
  \centerline{\epsfxsize=0.45\textwidth\epsffile{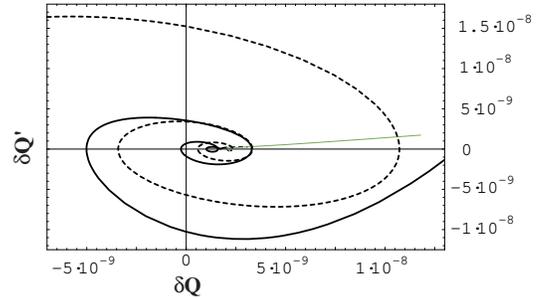}}
  \caption{Phase diagram for quintessence fluctuations with 
two different initial conditions. The attractor point
$(-2 \Phi V_{,Q}/V_{,QQ} , 0)$ is a 
transient attractor, valid only for long wavelength modes. As
soon as the tracking regime of the field $Q$ terminates, the attractor
disappears and the field perturbations start to evolve (green straight
line.) However, by the time tracking is over, most long-wavelength 
solutions have already converged to the same value, and thereafter 
their evolution is almost indistinguishable.}
%The only fixed point of the map is $(0,0)$ which is out of the plot 
%and it will be reached only after the modes have become sub-Hubble.}
\label{fig3}
\end{figure}

%Notice that if one sets the initial conditions
%at a time when quintessence is comparable to the radiation, then
%the initial field perturbations $\delta Q_i$ and 
%$\delta \dot{Q}_i$ can affect the value at which the newtonian potential
%settles down, $\Phi^c$. In this sense, changing the initial 
%conditions can change the numerical value of the attractor solution.

Summarizing the results of this section: 
in the kinetic and potential phases there is a constant 
long wavelength solution which is a linear combination of 
the initial field fluctuations and of the gravitational 
potential, as one can see from the initial evolution in Fig. 2A. 
In the tracking period the quintessence perturbation  
stabilizes at the attractor solution (\ref{ndec})
and it remains at that constant value until the perturbation
reenters the Hubble radius, or until quintessence starts to dominate 
the background.

\section{Evolution of isocurvature perturbations}

Cosmological fluctuations are often characterized in terms of
the gauge-invariant curvature perturbation on comoving 
hypersurfaces, defined as \cite{MFB}:
\be
\label{def:zeta}
\zeta \equiv \frac{2}{3}\frac{\dot{\Phi}/H + \Phi}{ 1 + w} + \Phi \; .
\ee
The time variation of the intrinsic curvature
$\zeta$ is given, on large scales, by the non-adiabatic pressure or,
equivalently, by the amplitude of the isocurvature perturbations --- 
see Eq. (\ref{zeta}). 
Therefore, if the non-adiabatic pressure vanishes then 
$\zeta$ is constant.

The pressure perturbations can be split into adiabatic and
non-adiabatic components:
\be
\label{dP}
\delta p = \left. \frac{\delta p}{\delta \rho} \right|_{\delta \Gamma=0}
\delta \rho + \left. \frac{\delta p}{\delta \Gamma} \right|_{\delta \rho =0}
\delta \Gamma \; ,
\ee
where we define the entropy perturbation $\delta \Gamma$ as \cite{Wands}:
\be
\label{Gamma}
\delta \Gamma \equiv \frac{\delta p}{\dot{p}}-\frac{\delta \rho}{\dot{\rho}} 
\; .
\ee
As a consequence we have that the adiabatic pressure perturbation
is given by
\be
\label{soundv}
\delta p_{\rm ad} = c_s^2 \delta \rho \equiv
\left. \frac{\delta p}{\delta \rho} \right|_{\delta \Gamma=0} \delta \rho
= \frac{\dot{p}}{\dot{\rho}} \delta \rho \; ,
\ee
and the non-adiabatic pressure is given by the second 
term in the right-hand side of the definition (\ref{dP}):
\begin{eqnarray}
\label{dpnad}
\delta p_{\rm nad} &=& 
\left. \frac{\delta p}{\delta \Gamma} \right|_{\delta \rho =0}
\delta \Gamma =  \dot{p} \, \delta \Gamma
\\ \nonumber
&=& \delta p - c_s^2 \delta \rho 
\equiv \sum_i \left( \delta p_i - c_s^2 \delta \rho_i \right) \; .
\end{eqnarray}

We now give an analytic description of the time evolution of isocurvature
perturbations for long wavelengths. We work perturbatively assuming 
$\zeta$ constant, and
compute the non adiabatic pressure $\dot{p} \, \delta \Gamma$: when the
integrated effect is large, it means that isocurvature perturbations
cannot be neglected. The results
of this analysis confirm the methods of the 
previous section (where $\Phi$ was assumed constant) 
and are in agreement with the numerical analysis.

Using Eq. (\ref{dpnad}) we find that 
the non-adiabatic pressure is given by:
\be
\label{dpnad2}
\delta p_{\rm nad} = \left( w_r - c_s^2 \right) \delta \rho_r 
+ \delta p_Q - c_s^2 \delta \rho_Q 
\; ,
\ee
By using Eqs. (\ref{wOmega}) and (\ref{cs2Om}) into
Eqs. (\ref{dpnad2}) one can check that in general 
\be
\label{dzeta}
\frac{\delta p_{\rm nad}}{\rho + p} =
{\cal O}(\Omega_Q) \times (\delta_r + \delta_Q) \; .
\ee
Therefore, when the quintessence contribution to the total energy
density is very subdominant, the isocurvature
contribution is small. 
However, since the isocurvature contribution to $\zeta$ is an
integrated effect, $\dot \zeta \sim t \delta p_{\rm nad}$, we 
should study
the time evolution of each term which enters into the definition of
the non-adiabatic pressure (\ref{dpnad2}).

The first term in the right-hand side of Eq. (\ref{dpnad2}) is 
proportional to $\rho_r \delta_r$ and leads at most to a 
logarithmic increase of $\zeta$.

Among the contributions from the pressure and the energy density of the
scalar field we neglect the first term in the right-hand sides of
both Eqs. (\ref{drQ}) and
(\ref{dpQ}). The term $\dot{Q}^2 \Phi$ is suppressed during the kinetic
phase, but leads to a growth as $a^{7 + 3w_{\rm TOT}}$ in the
left-hand-side of Eq. (\ref{dzeta}) in the potential regime.
During the tracking regime $\dot{Q}^2$ gives approximately a constant
contribution to (\ref{dzeta}).
The term $V,_{Q} \delta Q$ depends explicitly on the quintessence
fluctuations: it decays during the kinetic phase and it grows 
less rapidly than $\dot{Q}^2 \Phi$ during the potential regime. 
However, it leads to
a growth as $a^{3/2(1+w_{\rm TOT})}$ during the tracking regime.

It is clear from Eq. (\ref{dpnad}) that for two barotropic fluids with the
same equation of state and the same density contrast the non-adiabatic
pressure should be zero. Therefore for {\em exact} tracking, quintessence
and radiation equilibrate to give zero non-adiabatic pressure. We can also
compute the non-adiabatic pressure in the tracking regime by using Eqs.
(\ref{dpnad2}), (4) and (\ref{gamma_appr}): 
\begin{eqnarray}
\delta p_{\rm nad} & \simeq &
\frac{\dot{Q}^2}{\rho + p} (w_r - c_Q^2) \left( \rho_r \delta_r - 2 \Phi
\rho_Q \right) \\ \nonumber
& & + {\cal O} [ (\gamma -1) \rho_Q \Phi] 
\end{eqnarray} 
%It is easy to check
%that for exact tracking $c_s^2 =w=w_Q$ and hence $\delta p_{\rm nad}=0$.
As expected, the non-adiabatic pressure vanishes for exact tracking
($\gamma = 1$), otherwise it is small, but not vanishing, during tracking. In
general $\delta p_{\rm nad}$ is proportional to $\rho_Q$, therefore small
in many models. 

%Therefore, it seems difficult to generate and sustain
%isocurvature fluctuations at early times in quintessence models.
%For models without tracking, this is
%because quintessence is unimportant before very late times. 
%For models with tracking, quintessence may be a non-negligible
%component, but it behaves like the background fluid and hence
%its perturbations are indistinguishable from the matter perturbations.

There is however still one possibility that allows for significant
isocurvature fluctuations from quintessence: this happens
when the tracking phase starts only at a relatively late redshift, 
$z\sim 10^5 - 10^3$ (see also the next Section, Figs. 4, 5 and 6.) 
During the transition from the potential
phase to the tracking phase, we can have {\em both} $\Omega_Q$ non-negligible,
{\em and} the equation of state and speed of sound of
quintessence differ substantially from those of the background fluid.
If the quintessence density contrast 
is of the same order of the matter density contrast at decoupling time 
$z_{\rm dec}\sim 10^3$, 
the isocurvature perturbations can leave an imprint on 
the CMBR. 
%via the Sachs-Wolfe and early integrated Sachs-Wolfe effects.
Later, as tracking forces the field perturbations to the attractor, the
isocurvature fluctuations are temporarily depressed, at least until 
$Q$ starts to dominate at $z\sim 0$.

\section{Initial conditions: adiabatic or mixed?}

In this section we address the issue of how initial conditions of
quintessence perturbations can be used as tools in CMBFAST \cite{SZ}.
These initial conditions are set after nucleosynthesis, 
at $z \sim 10^9$. In most of the literature, the initial conditions for 
the quintessence fluctuations are set up by requiring adiabaticity with
the other components \cite{VL}. However,
the notion of a purely adiabatic perturbation (as well for a purely
isocurvature one) is an instantaneous notion for a multifluid system.
Moreover, because of the unthermalized nature of quintessence, 
the adiabatic condition for this component is even less justified.

The adiabatic condition\cite{VL,PB} is usually defined as
the vanishing of the relative entropy and its time derivative:
\begin{eqnarray}
\label{S_ad}
{S}_{rQ} &=& 0 \; ,
\\ \label{Sdot_ad}
\dot{S}_{rQ} &=& 0 \; ,
\end{eqnarray}
which reduces, in longitudinal gauge, to:
\begin{eqnarray}
\label{adiabatic1}
\delta Q &=& \dot{Q}^2 \left( V,_Q 
- \frac{k^2}{a^2} \frac{\dot{Q}}{6 H} \right)^{-1}
\left( - \frac{\delta_r V,_Q}{4 H \dot{Q}} 
+ \frac{k V_r}{6 a H} \right) \; ,
\\ \label{adiabatic2} \delta \dot{Q} 
&=& \dot{Q}^2 \left( V,_Q - \frac{k^2}{a^2} \frac{\dot{Q}}{6H} \right)^{-1} 
\left[ - \frac{k^2}{6 a^2 H} \left( \frac{3}{4}
\delta_r + \Phi \right) \right.
\\ \nonumber
& & \left. + V,_Q \left( \Phi - \frac{\ddot{Q} \delta_r}{4 H \dot{Q}} - 
\frac{k V_r}{6 a H} \right) 
 \right] \; .
\end{eqnarray}
The relative entropy between the radiation
and the quintessence components is defined 
as\footnote{Notice that the definition (\ref{def_S}) of entropy 
is different from the $\delta \Gamma$ introduced in Eq. (\ref{Gamma}).
We consider this more standard definition $S_{rQ}$ as well
since it is this quantity which is used in much of the literature 
to define the adiabatic conditions, and we want to show directly 
the difference between the initial conditions 
(\ref{adiabatic1})-(\ref{adiabatic2}) and the ones
defined by the attractor or by some previous dynamics.}:
\be 
\label{def_S}
S_{rQ} \equiv \frac{\delta_r}{1 + w_r} - \frac{\delta_Q}{1 + w_Q} \,.
\ee
From the above relation we immediately understand that $S_{rQ}=0$ in the
case of {\em exact} tracking. Indeed, two fluids with the same equation
of state and the same density contrast are undistinguishable.
For long wavelengths relations (\ref{adiabatic1})-(\ref{adiabatic2})
reduce to:
\begin{eqnarray} 
\label{cond1}
\delta Q &\simeq& - \frac{\dot{Q}}{4 H} \delta_r \simeq
\frac{\dot{Q}}{2 H} \Phi 
\\ \nonumber
{\delta \dot{Q}} &\simeq& \dot Q \left[ \Phi + \frac{\delta_r}{4} 
\left( 3 + \frac{V,_Q}{H \dot{Q}} \right) \right] 
\\ \label{cond2}
&\simeq& - \frac{\dot{Q}}{2} \, \Phi \left( 1 + \frac{V,_Q}{H \dot{Q}}
\right)
\end{eqnarray}
where the second line of Eq. (\ref{cond2}) holds only if radiation is the 
dominant component.
%, and the last equality in the second line of 
%Eq. (\ref{cond2}) follows from Eq. (\ref{VpV}).

The conditions (\ref{cond1})-(\ref{cond2}) should be 
compared with the solutions during the kinetic and
potential phases, 
$\delta Q=$constant and ${\delta \dot{Q}}=0$,
or with the attractor solution Eq. (\ref{ndec}) 
which applies during the tracking phase:
\begin{eqnarray}
\label{att_iso1}
\delta Q &\simeq& - 2 \frac{V,_{Q}}{V,_{QQ}} \Phi 
\\ \label{att_iso2}
{\delta \dot{Q}} &\simeq& - 2 \dot{Q} \, \Phi \left( 1 - \frac{ V_{,Q}
V_{,QQQ}}{V^{2}_{,QQ}} \right) \; .
\end{eqnarray}
It is not hard to see from Eq. (\ref{att_iso2}) 
that in the tracking regime $\delta \dot{Q} \rightarrow 0$, 
since $V_{,Q}/V_{,QQ} \simeq V_{,QQ}/V_{,QQQ}$. Using this fact
together with Eq. (\ref{VpV}), we also obtain that Eqs. (\ref{cond1})
and (\ref{att_iso1}) are similar during the tracking regime. Therefore,
as expected, if the scalar field is tracking normal matter, then
the adiabatic condition is approximately satisfied.
%isocurvature perturbations are suppressed in amplitude during tracking.

We illustrate the previous discussion in Figs. 4, 5 and 6. 
The background model is plotted in Fig. 4, and in Figs. 5-6
we present possible scenarios for the perturbations, as well
as a comparison with the cosmological perturbations in the
$\Lambda$CDM case. 

The upper (thin red) curves of Fig. 5 plot
the gauge-invariant curvature $\zeta$, defined in Eq. (\ref{def:zeta}).
The lower (black) curves are the Newtonian potential $\Phi$.
All plots in Fig. 5 have been normalized so that $\Phi = 5. 10^{-6}$ at 
$z=10^9$, and the wavelengths corresponds to modes which are crossing
the Hubble radius at the present time ($z=0$.)

\begin{figure}
\centerline{\epsfxsize=0.45\textwidth\epsffile{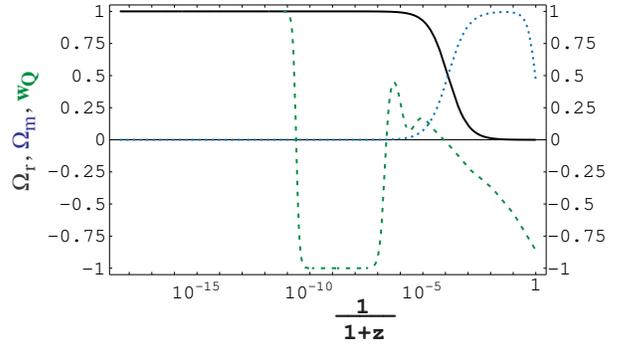}}
\caption{Background evolution in a quintessence scenario with the
Ratra-Peebles potential $V(Q)=M^{4+\alpha} Q^{-\alpha}$, where 
$\alpha=6$ and $M=10^{-6.15}$. Plotted are the densities of 
radiation (solid line, black) and matter (dotted
line, blue), and the equation of state for the scalar field (dashed line,
green), as a function of redshift.}
\label{5.1}
\end{figure}

The solid lines of Fig. 5 correspond to the cosmological perturbations of
a fiducial $\Lambda$CDM scenario with adiabatic initial conditions.
Notice that, as usual, since $\zeta$ remains constant, $\Phi$ has to change 
by $9/10$ after $z_{eq}=10^4$. 

The long-dashed lines of Fig. 5 are the perturbations in
a scenario (ICDM) where, in addition to the adiabatic mode, CDM and 
radiation have an initial relative isocurvature of $S_{rCDM}=3\Phi/2$.

The short-dashed lines are the perturbations in the case of
adiabatic initial conditions (AIC) between all components at $z
\sim 10^{9}$.
The dotted lines are the perturbations in the case where we
choose zero isocurvature between radiation and CDM, and $\delta Q =
10^{-3}$,
$\delta \dot{Q}=0$ initially (QIC.)
This last set of initial conditions (QIC) is motivated by the fact 
that the field perturbations are constant during the kinetic 
and the potential phases --- see, e.g., Fig. 2A.

Notice the identical late isocurvature effect in AIC and QIC.
The signal of this effect is the extra growth of the Bardeen 
parameter $\zeta$ at late times (compare the dotted and short-dashed
lines with the solid line in the interval $0<z<10$ in Fig. 5.)
This effect is independent from the initial conditions 
for the quintessence fluctuations.
In fact, as already emphasized, the notion of adiabaticity
(as well as pure isocurvature) is an instantaneous one, imposed at
an initial time, and it does not persist in a multi-fluid system. We stress
that this effect is distinct from the change of the Newtonian
potential $\Phi$ which is due to the late change in the equation of 
state of the background, which can be seen in pure form in the 
$\Lambda$CDM case (solid line in Fig. 5.)
A similar change in $\Phi$ occurs also at $z_{\rm eq}=10^4$, while $\zeta$ 
remains constant across the transition between radiation- and
matter-domination.

%??? Due to the different time evolution of radiation, CDM, and $Q$
%sub-Hubble perturbations, amounts of isocurvature CDM-radiation have a
%very distinctive effect \cite{EB,PIER,ENKV}. ????

\begin{figure}
\centerline{\epsfxsize=0.45\textwidth\epsffile{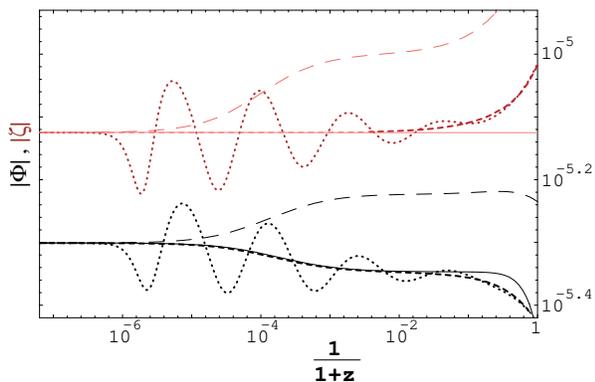}}
\caption{Newtonian potential (lower, black curves) and gauge-invariant
curvature $\zeta$ (upper, red curves), normalized by the
condition that $\Phi = 5. 10^{-6}$ at $z=10^{9}$. 
The solid lines are a fiducial $\Lambda$CDM model. The remaining lines
correspond to different initial conditions for the fluctuations of the
radiation, matter and quintessence components. The long-dashed lines
correspond to the case where there is no isocurvature component 
between $Q$ and radiation,
but there is an initial isocurvature component between radiation and CDM.
The short-dashed line corresponds to the case of pure adiabatic initial 
conditions (AIC). The dotted lines correspond to the 
case of $\delta Q=10^{-3}$, $\delta \dot{Q}=0$ initially (QIC). 
All modes cross the Hubble radius at $z=0$.}
\label{5.2}
\end{figure}

\begin{figure}
\centerline{\epsfxsize=0.45\textwidth\epsffile{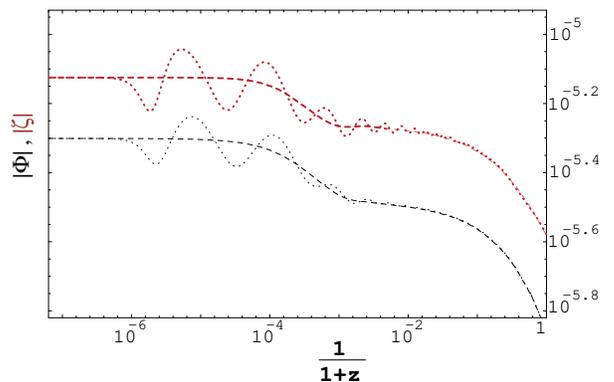}}
\caption{Newtonian potential (lower, black curves) and gauge-invariant
curvature $\zeta$ (upper, red curves), for a wavelength which crosses the
Hubble radius at decoupling time between matter and radiation.
The normalizations and the initial conditions are the same of Fig. 5.
As in Fig. 5, the short-dashed lines correspond to the case of pure
adiabatic initial conditions and the dotted lines correspond to the case
of $\delta Q=10^{-3}$, $\delta \dot{Q}=0$ initially. Even for
smaller scales, the effect of an isocurvature component is a transient
effect. For wavelengths larger than the Hubble radius, the attractor was
responsible for explaining the transiency, while for smaller scales the
explanation is the decay of quintessence fluctuations inside the Hubble
radius.}
\label{fig6}
\end{figure}

Notice also the early isocurvature effect in the QIC case (oscillations
of the dotted lines in Fig. 5.) 
This means that there is a substantial non-adiabatic pressure,
and hence a large isocurvature perturbation, in this scenario. As we 
discussed at the end of the previous Section, the reason 
why isocurvature fluctuations can become important
is that the ``tracking regime'' of the background model only starts 
relatively late, at $z \sim 10^5$. But the quintessence field 
perturbations need some time to converge to the attractor solution. 
During this time $\Omega_Q$ becomes non-negligible, and since
the $Q$ component still behaves quite differently from the dominant
background fluid, there can be substantial isocurvature perturbations for 
wavelengths which cross the Hubble radius at $z \sim 0$.
A similar transient effect occurs also for smaller scales, as shown 
in Fig. 6, for a wavelength which crosses the Hubble radius at the
decoupling time. An isocurvature component from the quintessence field
leads only to a transient effect, since quintessence fluctuations
converge to an attractor for long wavelengths, and decay in
time inside the Hubble radius. 
However, this early transient isocurvature effect can leave an imprint in
the spectrum of the CMBR anisotropies.
 
%through the Sachs-Wolfe and early integrated Sachs-Wolfe effects,
%even if the isocurvature fluctuations are suppressed after the time
%of decoupling, $z_{\rm dec}=10^3$. For 

\section{Discussion and Conclusions}

We have studied the evolution of cosmological perturbations
in quintessence models, with particular attention paid to isocurvature
modes generated by quintessence. We have shown that these isocurvature
modes are generic, with the exception of the case in which quintessence 
mimics {\em exactly} radiation. This occurs in the tracking phase of 
models with exponential potentials \cite{FJ}. 
However, these models are unappealing from the phenomenological point of
view because their equation of state for quintessence is not negative at
the present time.

When tracking is {\em not} exact, then isocurvature modes are
non-vanishing.  We found an {\em attractor} solution for long wavelength
quintessence fluctuations in the tracking regime. This allows an 
estimation of the amount of isocurvature fluctuations in the tracking 
regime for any quintessence model that display a tracking period. 
In the other phases which usually occur in models
with quintessence --- the kinetic and potential phases --- the
quintessence fluctuations have also a constant mode, whose actual value 
is determined by the previous evolution (including inflation.) 
The contribution of isocurvature fluctuation to the adiabatic 
mode grows during the potential, tracking and $Q$-domination phases.

We have discussed the assumption of adiabaticity in the context of the 
setting of initial conditions used in numerical codes such as CMBFAST. 
As already emphasized, because of the lack of 
thermal equilibrium, this assumption is not justified for 
quintessence. However, in the tracking case, the conditions set 
by the attractor solution are in most cases quantitatively 
indistinguishable from the adiabatic conditions. 
Indeed, tracking seems a gravitational mechanism --- alternative 
to thermal equilibrium --- which tends to reduce isocurvature modes 
between quintessence and the background fluid. 
For models with a tracking phase which starts early in time, we expect 
a weak dependence on the initial conditions for the
quintessence fluctuations. 
In the stages prior to the tracking phase the ``initial 
conditions'' (understood as the values of the field perturbation
and its time derivative at a redshift of $z \sim 10^9$) 
depend on the conditions set by inflation, and could be 
different from the adiabatic ones.

We have identified a late isocurvature effect for long wavelengths 
due to quintessence. Of course, this is due to the fact that Q 
dominates at late times, and it is qualitatively independent 
from the model considered. Therefore, the theoretical explanation 
of the long wavelength evolution of the Newtonian potential in 
Q models is a superposition of two effects: the change of the 
equation of state and the growth of $\zeta$ on super-Hubble scales.
This late isocurvature effect could be useful in order to distinguish
a cosmological constant model from the models with a quintessence component.

The observational relevance of isocurvature modes generated by 
quintessence is weakened by the decay in time of quintessence 
perturbations inside the Hubble radius. For this reason the 
isocurvature mode in the Q-radiation sector are very different 
from those in CDM-radiation.
This effect was also very appealing in order to minimize the effects
of the inclusion of this extra component on structure formation.
In the models examined here, the suppression of $\Omega_Q$ 
during the kinetic phase play also a crucial role in order to 
weaken the effect of some initial isocurvature modes generated by
quintessence. Even if the upper bound during 
nucleosynthesis for $\Omega_Q$ is $\sim 0.2$ ,
the kinetic regime suppresses $\Omega_Q$ down to $10^{-15}-10^{-20}$ 
in the models which we have analyzed. Therefore our considerations 
may be more relevant for models in which $\Omega_Q$ is closer 
to the upper limit, as, for instance, in the models with a modified
exponential potential \cite{AS}.

It is therefore interesting to study 
the impact of isocurvature fluctuations in quintessence 
models using numerical tools such as CMBFAST \cite{SZ} 
in which quintessence perturbations are included. 
In particular, it is possible to construct models where 
$\Omega_Q$ is not so suppressed or in which there is a late 
tracking phase, in which case isocurvature modes generated by 
quintessence could lead to observable effects in the temperature 
anisotropies of the cosmic background radiation.

\vskip 0.2cm

{\bf Acknowledgments}

\noindent
We would like to thank C. Baccigalupi, R. Brandenberger, V. Mukhanov 
and N. Turok for comments.
R. A. thanks the Physics Department at Purdue University for its 
hospitality at the time of writing the first draft of this paper.
F. F. also thanks the Ludwig Maximilians Universit\"at for its 
hospitality during the final stages of this work.
This work was supported in part by the U.S. Department of Energy
under Grant DE-FG02-91ER40681 (Task B), and by the 
Sonderforschungsbereich 375-95 f\"ur Astro-Teilchenphysik 
der Deutschen Forschungsgemeinschaft.

\section{Appendix A}

Here we present the equations that were evolved numerically.
Combining the $0-0$ and $i-i$ Einstein field equations we can eliminate
the radiation perturbations, and obtain the following equation for
the quintessence and matter perturbations (as explained in the text, we
work with units such that $8 \pi G = 1$, and we have ignored the 
distinction between baryons and CDM):
\begin{eqnarray}
\label{num1}
3 \ddot\Phi &+& 15 H \dot\Phi + \left( 12H^2 + 
\frac{k^2}{a^2} \right) \Phi
\\ \nonumber
&& = -\frac12 \delta_m \rho_m + \dot{Q}\delta\dot{Q} - \dot{Q}^2 \Phi 
-2 V_{,Q} \delta Q \; .
\end{eqnarray}
The equation of motion for the scalar field 
perturbations is given in Eq. (\ref{fieldpert}).
The matter density contrast obeys the energy conservation equation:
\be
\label{num2}
\dot{\delta}_m - 3 \dot \Phi - \frac{k}{a} v_m = 0 \; ,
\ee
where $v_m$ is the matter fluid velocity, which itself satisfies
the momentum conservation equation:
\be
\label{num3}
\dot{v}_m + H v_m + \frac{k}{a} \Phi = 0 \; .
\ee

\section{Appendix B}

Here we present the evolution of pure isocurvature modes generated by
the quintessence sector. Consider contributions from
quintessence and another component $X$ to the metric perturbation,
which cause a vanishing contribution to $\Phi$ and $\dot \Phi$ \cite{MFB}
at some fixed initial time in the energy and momentum constraint. 
This requires that:
\begin{eqnarray}
\Omega_Q \delta_Q = - \Omega_X \delta_X
\label{isoQX_en}
\\
\dot{Q} \delta Q = - \frac{4}{3} v_X \rho_X \,,
\label{isoQX_v}
\end{eqnarray}
which identifies an isocurvature density mode (\ref{isoQX_en}) and an
isocurvature velocity mode (\ref{isoQX_v}) \cite{BKT}.
The third component $Y$ is the only one which leads to a curvature
perturbation (for long wavelengths $-2 \Phi = \Gamma_Y \delta_Y$, as
follows from Eq. (\ref{EFEapp}).) The components $X$ and $Y$ are related
by adiabaticity ($S_{XY} = 0$.) Therefore, the quintessence density
contrast at fixed initial time is:
\be
\delta_Q = - \delta_X \frac{\Omega_X}{\Omega_Q} = 2 \Phi \left( \frac{1 +
w_X}{1+w_Y} \right) \frac{\Omega_X}{\Omega_Y \Omega_Q} 
\label{delta_iso}
\ee   

In Fig. 7 the evolution of the Bardeen parameter $\zeta$ for initial pure
Q-CDM ($X={\rm CDM}$, solid line) and 
Q-radiation ($X={\rm radiation}$, dashed
line) isocurvature initial
conditions at $z = 10^9$ for the model of Fig. 4. 
At the initial time the Newtonian potential is so small in order to ensure 
$\delta_Q \le 1$. However, in both cases - but in particular for the
Q-radiation isocurvature initial conditions -, Eqs. 
(\ref{isoQX_en})-(\ref{isoQX_v}) 
force Q fluctuations to very large values, either initially
and in the subsequent evolution. The evolution of Q fluctuations for these
type of initial conditions is {\em not} described by the attractor
(\ref{ndec}) -- i.e. the inhomogeneous solution to the differential
equation for the quintessence fluctuations --,
but by the homogeneous solution to Eq. (\ref{fieldpert}).  
Indeed, in Fig. 7 the growth of the Bardeen parameter saturates
roughly when tracking starts in Fig. 4, i.e. when the homogeneous solution
for quintessence fluctuations start to decay, as described by Eqs. 
(19)-(20) with $\alpha \ne 0$.

\begin{figure}
\centerline{\epsfxsize=0.45\textwidth\epsffile{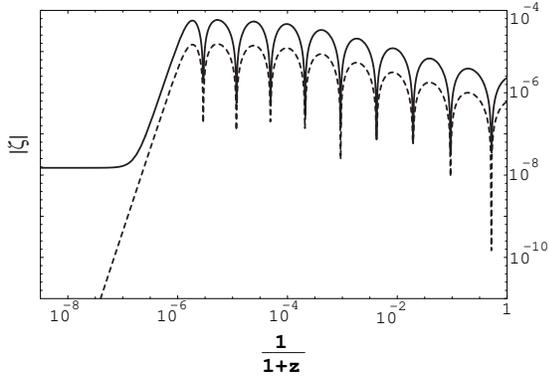}}
\caption{Gauge-invariant curvature $\zeta$ for Q-CDM isocurvature (solid
lines) and Q-radiation (dashed lines)
initial conditions for a
wavelength which crosses the Hubble radius at $z=0$.
The normalization is made in such a way that the amplitude of the
Newtonian potential is roughly $10^{-5}$ when the growth saturates.}
\label{fig7}
\end{figure}

%We now discuss isocurvature initial conditions generated by the
%quintessence component. Consider contributions from
%quintessence and another component $F$ to the metric perturbation,
%which cause $\Phi = \dot \Phi =0$ \cite{MFB} at some fixed initial
%time. This requires that:
%\begin{eqnarray}
%\Omega_Q \delta_Q = - \Omega_F \delta_F
%\label{isoQX_en}
%\\
%\dot{Q} \delta Q = - \frac{4}{3} v_F \rho_F \,.
%\label{isoQX_v}
%\end{eqnarray}
%If one considers $F=r$ then establishing isocurvature
%conditions between quintessence and radiation could be problematic.
%In fact, if radiation dominates ($\Omega_r \simeq 1$) one
%has, from Eq. (\ref{EFEapp}):
%\be
%\delta_Q \simeq \frac{2 \Phi}{\Omega_Q}
%\ee
%leading to $\delta_Q \gg 1$ when $\Omega_Q \ll 10^{-5}$.
%Isocurvature modes between $Q$ and CDM, on the other hand,
%can be realized in the context of
%linear theory, since $\Omega_{m} \ll 1$. 

\end{document}